\documentclass[preprint,aps,superscript,address,showkeys,showpacs,pre]{revtex4}
\usepackage{amsmath,amssymb}
\usepackage{epsfig}
\usepackage{enumerate}
\usepackage{graphics}
\usepackage{graphicx}
\usepackage{float}
\usepackage{dcolumn}
\begin{document}

\title{Signal amplification by unidirectional coupling of oscillators}

\author{S.~Rajamani}
\email{rajeebard@gmail.com}
\author{S.~Rajasekar}
\email{rajasekar@cnld.bdu.ac.in, srj.bdu@gmail.com}
\affiliation{School  of Physics, Bharathidasan University, Tiruchirapalli 620 024, Tamilnadu, India}

\begin{abstract}
We report our investigation on the input signal amplification in unidirectionally coupled monostable Duffing oscillators in one- and two-dimensions with first oscillator alone driven by a weak periodic signal. Applying a perturbation theory we obtain a set of nonlinear equations for the response amplitude of the coupled oscillators. We identify the conditions for undamped signal propagation with enhanced amplitude through the coupled oscillators. When the number of oscillators increases the response amplitude approaches a limiting value. We determine this limiting value. Also, we analyse the signal amplification in the coupled oscillators in two-dimensions with fraction of oscillators chosen randomly for coupling and forcing.

\pacs{02.30.Mv, 05.45.Xt, 46.40.Ff, 05.45.-a.}  
\keywords{Coupled Duffing oscillators, unidirectional coupling, signal amplification, resonance.}
\end{abstract}
\maketitle

\newpage

\section{Introduction}

Over the past three decades or so a great deal of interest has been paid on exploring the various features of nonlinear systems, particularly, in systems driven by a periodic force. Often a periodic force is treated as a simple external perturbation to induce oscillatory motion and different nonlinear phenomena in damped nonlinear oscillators. In many physical, engineering and biological systems and in electronic circuits it is not only easy to introduce a periodic force but it also represent an inherent part of them. For example, a periodic force represents a periodic seasonal variation in ecological systems {\cite{r1}}, an injection current in lasers {\cite{r2}}, imaging and audio pumping waves in an acoustic field {\cite{r3}}, oscillations introduced to mechanical devices by means of electromagnetic vibration generator {\cite{r4}}, an input voltage/current source in electronic circuits {\cite{r4,r5}}, membrane current in neuronal axons {\cite{r6,r7}}, pacemakers' activity in real life systems {\cite{r8}}, an applied illumination in photolithography-assisted techniques in chemical reactions {\cite{{r9},{r10}}} and so on.

Identification of a weak input periodic signal and enhancement of response of nonlinear systems are of important. We note that the response of a nonlinear system can be enhanced in different ways. For example, in nonlinear oscillators when the frequency $\omega$ of a periodic force is varied for a fixed small value of amplitude of the force, the response amplitude increases, reaches a maximum value at a particular frequency different from the natural frequency of oscillation of the systems and then decreases. This phenomenon is the well known nonlinear resonance. In bistable systems a weak  noise can improve the response at the frequency of the input signal. This phenomenon is called stochastic resonance {\cite{r11}}. It has been shown that resonance can occur in the absence of weak periodic force and is termed as coherence resonance {\cite{r12,r13}}. That is, noise alone is able to induce or enhance temporal order in the dynamics of certain nonlinear systems, particularly, in excitable systems. Apart from Gaussian white noise, resonance has been realized due to coloured noise {\cite{r14}} and chaotic signal {\cite{r15,r16}}. Noise-supported wave propagation in the experiments of photosensitive Belousov--Zhabotinsky chemical reaction {\cite{r17}}, one- and two-dimensional arrays of two-way coupled  bistable oscillators {\cite{r18}}, nonlinear cells with two-state threshold elements {\cite{r19}} and bistable oscillators driven by a biharmonic signal {\cite{r20}} and fault-tolerant behaviour in a chain of one-way coupled bistable oscillators {\cite{r21}} have been realized.  On the other hand, in the vibrational resonance approach the noise is replaced by a high-frequency periodic force {\cite{r22}}.

It is possible to enhance the response amplitude by unidirectionally coupling the oscillators. There are few interesting dynamics realized as a result of unidirectional coupling. One-way coupling was introduced by Visarath In and his collaborators {\cite{{r23},{r24}} to induce oscillations in undriven, overdamped and bistable systems. The mechanism for the generation of oscillation is described for a three-coupled overdamped Duffing oscillators {\cite{r25}}. The response of three-(one-way) coupled core magnetometer systems to a periodic magnetic-flux signal applied to all the three units is explored. The response is found to be either synchronized to the signal frequency or to one-third of it {\cite{r26}}. In one-way coupled systems propagation of waves of dislocations in equilibria are found {\cite{r27}}. One-way coupling is utilized in electronic sensors and microelectronic circuits {\cite{r28}} and is found to assist the propagation of localized nonlinear waves {\cite{r29,r30}}. Experimental evidences of propagation and annihilation of solitons in a mechanical array of unidirectionally coupled oscillators {\cite{r31}} and an electronic circuit in two-dimensions {\cite{r32}} are reported. Improved transmission of low-frequency signal by the combinative action of high-frequency input signal and one-way coupling has been found in coupled overdamped bistable oscillators {\cite{r33}} and in coupled maps {\cite{r34}}.

In biology also unidirectionally coupled structures constitute one of the simplest topologies of practical interest.  We wish to cite an example from the field of gene regulatory network.  A network of transcriptional regulators of considerable interest is the repressilator model.  It is the network consisting of $n$ genes with the protein product of $i$th gene represses the expression of $(i+1)$th gene with periodic boundary conditions.  Oscillatory dynamics is found in an experimentally constructed repressilator with three cells {\cite{r35}}.  Occurrence of monostable, spiral and limit cycle dynamics are analysed in a mathematical model of repressilator  with three genes {\cite{r36,r37}}.  A Repressilator system with even number of genes is found to exhibit multistability while with odd number of genes display a stable limit cycle {\cite{r38}}.  Existence of quasi-stable travelling wave periodic solutions is analysed both theoretically and numerically in unidirectionally coupled repressilators {\cite{r39,r40}}.

The goal of the present work is to investigate the enhancement of output signal amplitude and undamped signal propagation in unidirectionally coupled oscillators in one- and two-dimensions. In the coupled systems the first oscillator (unit) alone subject to a periodic force while the other oscillators are coupled unidirectionally. The coupling term is linear. For our analysis we choose the paradigm Duffing oscillator as the reference oscillator. For the coupled oscillators in one-dimension applying a perturbation theory we obtain a set of nonlinear equations for the amplitudes of the periodic oscillations of the oscillators. Below a critical value of the coupling strength $\delta$ the amplitude $A_i$ of $i$th oscillator decays to zero as $i$ increases even though $A_1$ of the first oscillator is nonzero. This corresponds to a damped signal propagation. When $A_i=0$ the state variables of the $i$th oscillator decays to zero as time $t \to \infty$.   Above the critical value of $\delta$, the response amplitude $A_i$ increases with $i$ and attains a saturation leading to an enhanced signal propagation through the coupled oscillators.  In this case $A_i > A_1$ for $i>1$ and the state variables of the oscillators not decays to zero with time but the oscillators exhibit periodic motion with the frequency $\omega$ of the periodic force applied to the first oscillator.  We refer  this case as an undamped signal propagation. From the nonlinear equations of the amplitudes $A_i$'s we are able to determine the saturate value of response amplitude and the threshold condition on $\delta$ for undamped signal propagation. The theoretical predictions agree quite well with the numerical simulation. For a fixed value of $\delta$ when the frequency $\omega$ of the external periodic force is varied the oscillators exhibit nonlinear resonance and hysteresis. The width of the hysteresis decreases with increases in the value of $\delta$ and then it disappears.

For the one-way coupled oscillators in two-dimensions with
\begin{enumerate}[(i)]
\item 
the first oscillator alone driven by the periodic force,
\item 
all the oscillators are driven but randomly chosen certain number of oscillators are alone coupled and 
\item 
all the oscillators are coupled but the periodic force is applied to randomly chosen certain number of oscillators only
\end{enumerate}
also the theoretical treatment enables us to determine the response amplitudes of the oscillators. Enhancement of response amplitude is realized in the above three cases.
\section{Coupled Duffing oscillators in one-dimension}
The equation of motion of the unidirectionally coupled $n$-Duffing oscillators of our interest in one-dimension is given by
\begin{subequations}
\label{eq1}
\begin{eqnarray}
 \ddot x_1 + d \dot x_1 + \omega_0^2 x_1 + \beta x_1^3
      & = & f \cos \omega t,  \label{eq1a} \\
 \ddot x_i + d \dot x_i + \omega_0^2 x_i + \beta x_i^3
      & = & \delta x_{i-1},   \quad i=2,3,\cdots,n.  \label{eq1b} 
\end{eqnarray}  
\end{subequations}
We assume that the values of all parameters in Eqs.~(\ref{eq1}) are $>0$. In {\cite{r26}} coupled magnetometer systems were considered wherein the units form a closed ring (periodic boundary conditions), the coupling is unidirectional and the signal was applied to all the units in the ring. In contrast to these, in the system (\ref{eq1}) the first oscillator alone driven by the periodic signal $f\cos\omega t$. The other oscillators are coupled unidirectionally. The dynamics of the first oscillator is independent of the dynamics of the other oscillators while that of the $i$th oscillator depends on the dynamics of $(i-1)$th oscillator through the linear coupling.  The last oscillator is not connected to the first oscillator.  The system (\ref{eq1}) is capable of showing a variety of nonlinear phenomena including chaotic dynamics. In our present work we are interested in resonance behaviour and enhancement of amplitude of periodic response of the system due to one-way coupling. Therefore, we choose $f \ll 1$ so that in Eq.~(\ref{eq1a}) periodic oscillation occurs about the equilibrium point.

Applying a perturbation theory to the system (\ref{eq1}) we obtain a set of coupled nonlinear equations for the amplitudes of the period-$T(=2 \pi / \omega)$ solutions of the oscillators and then analyse the influence of one-way coupling on response amplitudes and resonance.
\subsection{Theoretical treatment}
We assume the periodic solution of the system (\ref{eq1}) as 
\begin{equation}
 \label{eq2}
     x_i(t) = a_i(t) \cos \omega t + b_i(t) \sin \omega t,      
\end{equation}
where $a_i(t)$ and $b_i(t)$ are slowly varying functions of time $t$. We write
\begin{subequations}
 \label{eq3}
\begin{eqnarray} 
   x^3_i 
      & \approx & \frac{3}{4} \left( a_i^2 + b_i^2 \right)
                  \left( a_i \cos \omega t + b_i \sin\omega t \right), 
             \label{eq3a}  \\
     \dot x_i(t)
     & = & \dot a_i \cos \omega t + \dot b_i \sin\omega t
             - a _i \omega \sin \omega t + b_i \omega \cos\omega t,
                     \label{eq3b} \\     
   \ddot x_i(t)
     & = & - 2 \dot a_i \omega \sin \omega t 
               + 2 \dot b_i \omega \cos \omega t
               - a_i \omega^2 \cos \omega t
               - b_i \omega^2 \sin \omega t, \label{eq3c}
\end{eqnarray} 
\end{subequations}
where in Eq.~(\ref{eq3c}) we neglected $\ddot{a}_i$ and $\ddot{b}_i$ because of their smallness. Substituting (\ref{eq2})-(\ref{eq3}) in (\ref{eq1}), neglecting $d \dot{a}_i$ and $d \dot{b}_i$ as they are assumed to be small and then equating the coefficients of $\sin \omega t$ and $\cos \omega t$ separately to zero we obtain
\begin{subequations}
\label{eq4}
\begin{eqnarray}
    \dot a_i 
    & = & - \frac{b_i}{2\omega} \left[ \omega^2 - \omega_0^2
            - \frac{3}{4} \beta \left( a_i^2 + b_i^2 \right)\right]
            - \frac{1}{2} d a_i + S_a,  \label{eq4a} \\
    \dot b_i 
    & = & \frac{a_i}{2\omega} \left[ \omega^2 - \omega_0^2
           - \frac{3}{4} \beta \left(a_i^2 + b_i^2 \right) \right]
           - \frac{1}{2} d b_i + S_b, \label{eq4b}
\end{eqnarray}
\end{subequations}
where for $i=1$
\begin{subequations}
\label{eq5}
\begin{eqnarray}
     S_a = 0, \quad S_b = \frac{f}{2\omega}, \label{eq5a}
\end{eqnarray}
and for $i>1$
\begin{eqnarray}
    S_a = - \frac{\delta b_{i-1}}{2\omega}, \quad 
    S_b = \frac{\delta a_{i-1}}{2 \omega}. \label{eq5c}
\end{eqnarray}
\end{subequations}
Next, we introduce the transformation
\begin{equation}
\label{eq6}
    a_i(t) = A_i(t) \cos \theta_i(t), \quad 
    b_i(t) = A_i(t) \sin \theta_i(t)
\end{equation}
with $A_i^2 = a_i^2+b_i^2$. Then Eqs.~({\ref{eq4}}) become 
\begin{subequations}
\label{eq7}
\begin{eqnarray}
    \dot{A}_i 
    & = &  - \frac{1}{2} d A_i + S_A,  \label{eq7a} \\
    A_i \dot{\theta}_i
    & = & \frac{A_i}{2\omega} \left[ \omega^2 -\omega _0^2 
            - \frac{3}{4}\beta A_i^2 \right]
          + S_{\theta}, \label{eq7b}
\end{eqnarray}
\end{subequations}
where for $i=1$ 
\begin{subequations}
\label{eq8}
\begin{eqnarray}
    S_A  = \frac{f}{2\omega} \sin \theta_1, \quad 
              S_{\theta} = \frac{f}{2\omega} \cos\theta_1 \label{eq8a}
\end{eqnarray}
and for $i>1$
\begin{eqnarray}
    S_A 
    & = & \frac{\delta}{2\omega} A_{i-1} \sin (\theta_i 
             - \theta_{i-1}), \quad
    S_{\theta}
     =  \frac{\delta}{2\omega} A_{i-1} \cos(\theta_i 
           - \theta_{i-1}).    \label{eq8b}
\end{eqnarray}
\end{subequations}
The response of the system ({\ref{eq1}}) in the long time limit is periodic with period-$T(=2 \pi / \omega)$ provided $A_i(t)$ and $\theta_i(t)$ become constants as $t \rightarrow \infty $ and are the equilibrium points of Eqs.~({\ref{eq7}}). To find the equilibrium points of ({\ref{eq7}}) we set $\dot A_i = \dot{\theta}_i=0$, $ A_i(t)=A_i^*$,  $\theta _i(t)= \theta_i^*$ and drop $*$ in $A_i$ and $ \theta _i$.  We obtain 
\begin{equation}
\label{eq9}
    A_i^2 \left[ \omega_0^2 - \omega^2 + \frac{3}{4} \beta A_i^2\right]^2
         + d^2 \omega^2 A_i^2
       = \begin{cases}
             f^2 & {\mathrm{for}}  \; i= 1 \\
             \delta^2 A_{i-1}^2 & {\mathrm{for}} \; i>1
          \end{cases}
\end{equation}
and 
\begin{equation}
\label{eq10}
   \theta _i = \theta_{i-1} + \tan^{-1}
                \left[\frac{d \omega}{\omega_0^2 - \omega^2
                  + \frac{3}{4}\beta A_i^2} \right],
\end{equation}
where $\theta_0=0$.

When $i=1$ Eq.~(\ref{eq9}) is the frequency-response equation of the first oscillator. For $i>1$ Eq.~(\ref{eq9}) is independent of $f$ and depends on the coupling constant $\delta$ and the response amplitude $A_{i-1}$ of the $(i-1)$th oscillator. Equation (\ref{eq9}) (for $i>1$) is the coupling strength-response (amplitude) equation of the $i$th oscillator. The amplitude and phase of the $i$th oscillator with $i>1$ depend on the amplitude and phase respectively of the $(i-1)$th oscillator. $A_i$'s and $\theta_i$'s can be determined by solving the Eqs.~(\ref{eq9}) and (\ref{eq10}) successively.

\subsection{Analysis of effect of one-way coupling}
Now, we analyses the influences of the coupling strength on the response amplitudes $A_i$ and resonance. We fix the values of the parameters as $d=0.5$, $\omega_0^2=1$, $\beta=1$ and $f=0.1$. Equation (\ref{eq9}) contains even powers of $A_i$. It can be viewed as a cubic equation in terms of $A_i^2$. Explicit analytical expressions for the roots of a cubic equation is given in ref.\cite{r41}.  Equation (\ref{eq9}) can admit either one real root or three real roots. First, we determine the values of $A_1$ and then successively calculate $A_2$, $A_3$, $\cdots$. For our analysis we fix $n=200$.
 
In order to know the validity of the theoretical treatment we solve the Eqs.~({\ref{eq1}}) numerically using the fourth-order Runge-Kutta method. From the numerical solution we compute the amplitudes $A_i$. Figure {\ref{f1}} shows both theoretical and numerically computed $A_i$ versus $i$ for three fixed values of $\delta$ with $\omega=0.7$. 
\begin{figure}[b]
\begin{center}
\includegraphics[width=0.5\linewidth]{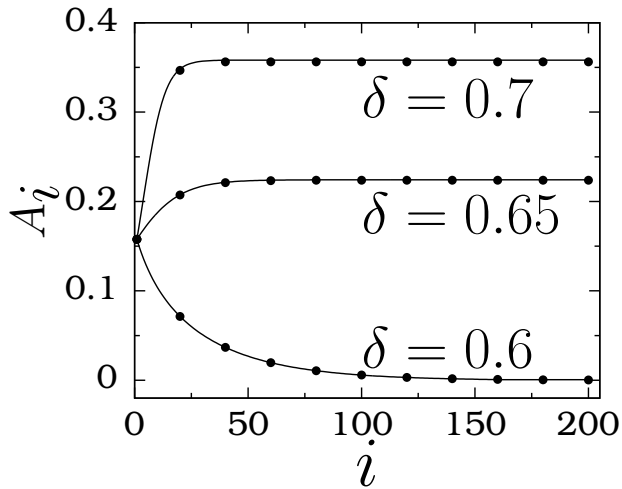}
\end{center}
\caption{Variation of the response amplitudes $A_i$ with $i$, the number of oscillators, for three values of the coupling strength $\delta$. The continuous lines and solid circles represent theoretical and numerical values of $A_{i}$. Numerical $A_i$'s are shown for certain selected values of $i$ for clarity. The values of the parameters are $d=0.5$, $\omega_0^2=1$, $\beta=1$, $f=0.1$ and $\omega=0.7$.}
\label{f1}
\end{figure} 
The theoretical $A_i$'s very closely match with numerically calculated $A_i$'s even for large values of $i$. For $\delta=0.6$ as $i$ increases the amplitudes $A_i$ of oscillation of the consecutive oscillators decrease and the distant oscillators do not oscillate but settle to the equilibrium state which is the origin. The coupled oscillators exhibit damped signal propagation ($A_i \rightarrow 0$ as $i\rightarrow \infty$). When $\delta=0.65$ and $0.7$ in Fig.~\ref{f1} we note that $A_i$ increases with $i$ and then reaches a saturation. Though the first oscillator alone is driven by the periodic force the one-way coupling gives rise enhanced signal propagation ($A_i > A_1$ for $i \gg 1$) over the coupled oscillators. This is a quintessence feature of one-way coupling. The point is that for the parametric values corresponding to the undamped signal propagation there exists a critical number of oscillators to realize a maximum value of the response amplitude of the last oscillator. This critical number of oscillators and the maximum amplitude depend on the control parameters $\omega$ and $\delta$.

Interestingly, the value of the saturate amplitude denoted as $A_{{\mathrm{L}}}$ (limiting value of amplitude in the limit of very large $i$) and the condition on $\delta$ for undamped signal propagation  can be obtained from the nonlinear equations of $A_i$ given by Eq.~(\ref{eq9}). $A_{{\mathrm{L}}}=0$ and $A_{{\mathrm{L}}} > A_1$ for damped and undamped signal propagation respectively. Equation  (\ref{eq9}) for $i>1$ can be viewed as a nonlinear map specifying the relation between $A_{i-1}$ and $A_i$. The stable equilibrium states of Eq.~(\ref{eq9}) are the values of $A_{{\mathrm{L}}}$. For $i>1$, the equilibrium states of Eq.~(\ref{eq9}) can be obtained by setting $A_i=A_{i-1}=A_{{\mathrm{L}}}$. This gives $A_{{\mathrm{L}}}=0$ and  
\begin{equation}
\label{eq11}
   A_{{\mathrm{L}}\pm}
     = \left[ \frac{4}{3\beta} \left( \omega^2 
          - \omega_0^2 \pm \sqrt{\delta^2 - d^2 \omega^2}
            \right) \right]^{1/2} .
\end{equation}

Now, we have two cases:

\noindent {\bf{Case $1$: $\omega^2 < \omega_0^2$ }}

When $\omega^2 < \omega_0^2$, $A_{{\mathrm{L}}}=0$ is the only equilibrium state for $\delta < \delta_{{\mathrm{c}}}$ where 
\begin{equation}
\label{eq12}
    \delta_{{\mathrm{c}}} 
    = \sqrt{\left( \omega^2 - \omega^2_0 \right)^2
        + d^2\omega^2} = \frac{f}{A_{{\mathrm{linear}}}}, 
\end{equation}
where $A_{{\mathrm{linear}}}$ is the response amplitude of the linear part of Eq.~({\ref{eq1a}}) $(\beta=0). $ $A_{{\mathrm{L}}}=0$ and $A_{{\mathrm{L}} \pm}$ are the equilibrium states for $\delta > \delta_{{\mathrm{c}}}$.

\noindent {\bf{Case $2$: $\omega^2 > \omega_0^2$ }}

$A_{{\mathrm{L}}}=0$ is the only equilibrium state for $\delta < d\omega$. For $d\omega < \delta < \delta_{{\mathrm{c}}}$ the equilibrium states are $A_{{\mathrm{L}}}=0$ and $A_{{\mathrm{L}}\pm}$. When $\delta > \delta_{{\mathrm{c}}}$ the possible equilibrium states are $A_{{\mathrm{L}}}=0$ and $A_{{\mathrm{L}}+}$. It is noteworthy to mention that $A_{{\mathrm{L}}\pm}$ are independent of the amplitude $f$ of the driving force but depend on the frequency of the driving force. They are inversely proportional to the coefficient $\beta$ of the nonlinear term. 

Next, we determine the condition on $\delta$ for enhanced and undamped signal propagation. The difference between the nonlinear equations for $A_1$ and $A_i$, $i > 1$ is only in the last term in Eq.~(\ref{eq9}). For $A_1$ the last term is $f^2$ while for $A_i$ it is $\delta^2 A^2_{i-1}$. Consider the Eq.~(\ref{eq9}) with $f^2$ (or $\delta^2A^2_{i-1}$) replaced by $\alpha$. Figure {\ref{f2}} shows the variation of the response amplitude $A$ ($A_1$ or $A_i$) for three fixed values of $\omega$. In all the cases $A$ increases monotonically with $\alpha$.   From Fig.~\ref{f2} we infer that in order to have $A_2 > A_1$ the condition is $\delta^2 A^2_1 > f^2$, that is,
\begin{equation}
\label{eq13}
   \delta > \delta_{{\mathrm{u}}} = \frac{f}{A_1}.
\end{equation}
For $\omega=0.7$ and $f=0.1$ we find $A_1=0.1577$ and $\delta_{{\mathrm{u}}}=0.634$. That is, undamped signal propagation occurs for $\delta > 0.634$ and is confirmed in the numerical simulation.

We point out the interplay between the parameters $d$ and $\delta$.  For fixed values of the parameters the amplitude $A_1$  is found to decrease with increase in the damping coefficient $d$.  Consequently, from Eq.~(\ref{eq13}) it is clear that $\delta_{{\mathrm{u}}}$ increases with increase in the value of $d$.  This is also evident from Eq.~(\ref{eq12}). $A_{{\mathrm{L}}\pm}$ given by Eq.~(\ref{eq11}) depend on all the parameters of the system except the amplitude $f$ of the input periodic force.  Further, $A_{{\mathrm{L}}+} ( A_{{\mathrm{L}}-})$ increases (decreases) with increase in the value of $\delta$.  In contrast to this, both $A_{{\mathrm{L}}+}$ and $A_{{\mathrm{L}}-}$ decrease with increase in $d$.  For $\omega=0.7$ and  $f=0.1$ the values of  $\delta_{{\mathrm{u}}}$ for $d=0.3$, $0.5$ and $0.7$ are $0.57275$, $0.63402$ and $0.71781$ respectively.  For $\delta=0.7$ the values of $A_{{\mathrm{L}}}$ for $d=0.3$, $0.5$ and $0.7$ are $0.45863$, $0.35818$ and $0$ respectively.  

\begin{figure}[t]
\begin{center}
\includegraphics[width=0.45\linewidth]{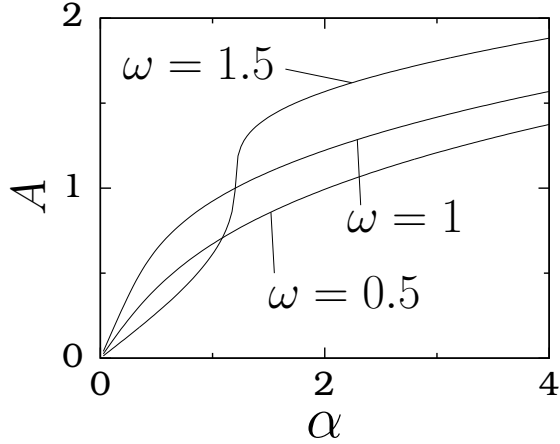}
\end{center}
\caption{Variation of the response amplitude $A$ ($A_1$ or $A_i$, $i>1$) with the parameter $\alpha$ ($=f^2$ for $A_1$ and $\delta^2 A_{i-1}$ for $A_i$) for three values of $\omega$.}
\label{f2}
\end{figure}
%
\begin{figure}[!h]
\begin{center}
\includegraphics[width=0.42\linewidth]{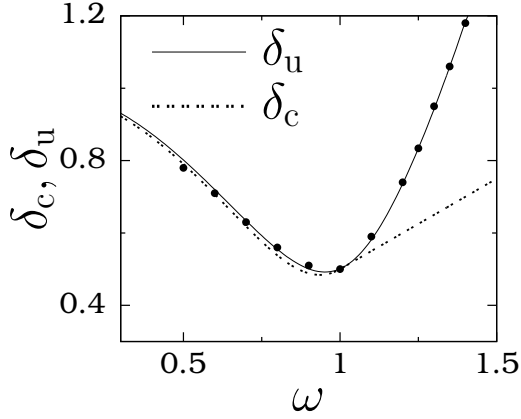}
\end{center}
\caption{Theoretical $\delta_{{\mathrm{c}}}$ (dotted curved) and $\delta_{{\mathrm{u}}}$ (continuous curve) given by Eqs.~(\ref{eq12}) and (\ref{eq13}) respectively versus $\omega$. The solid circles are the numerically computed values of $\delta_{{\mathrm{u}}}$.}
\label{f3}
\end{figure}

In Fig.~\ref{f3} we plot the dependence of $\delta_{{\mathrm{c}}}$ and $\delta_{{\mathrm{u}}}$ with $\omega$. For the values of $\omega$ and $\delta$ below the lower curve $A_{{\mathrm{L}}}=0$ is the only equilibrium state and is stable. In the regions between the upper and lower curves in addition to $A_{{\mathrm{L}}}=0$, a nontrivial $A_{{\mathrm{L}}}$ also exists with $A_{{\mathrm{L}}} (\neq 0) < A_1$. The nontrivial state is unstable while the state $A_{{\mathrm{L}}}=0$ is stable. In the regions above the upper curve $A_{{\mathrm{L}}}=0$ is unstable while the nontrivial $A_{{\mathrm{L}}} > A_1$ is stable leading to an enhanced undamped signal propagation. Undamped signal propagation with $A_{{\mathrm{L}}} (\neq 0) < A_1$ is not observed. Figure {\ref{f4}} presents $A_{{\mathrm{L}}}$ versus the control parameters $\delta$ and $\omega$. For each fixed value of $\omega$, $A_{{\mathrm{L}}}$ increases monotonically with $\delta$ for $\delta > \delta_{{\mathrm{u}}}$. 

\begin{figure}[t]
\begin{center}
\includegraphics[width=0.42\linewidth]{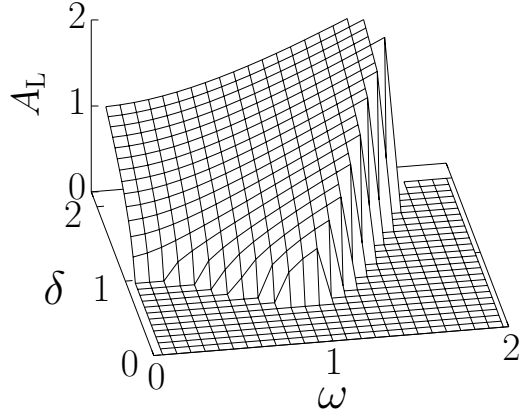}
\end{center}
\caption{Variation of the limiting value $A_{{\mathrm{L}}}$, the value of the response amplitude $A_i$ in the limit of $i \rightarrow \infty$, with the parameters $\delta$ and $\omega$. Zero and nonzero values of $A_{{\mathrm{L}}}$ represent damped and undamped signal propagations respectively.}
\label{f4}
\end{figure}

Figure {\ref{f5}} shows the frequency-response amplitude profile of various oscillators for a few fixed values of $\delta$. For $\delta=0.45$, in Fig.~\ref{f4} $A_{{\mathrm{L}}}=0$ for all values of $\omega$. In Fig.~\ref{f5}(a) for all values of $\omega$, $A_i$ decays to zero as $i$ increases. 
\begin{figure}[t]
\begin{center}
\includegraphics[width=0.5\linewidth]{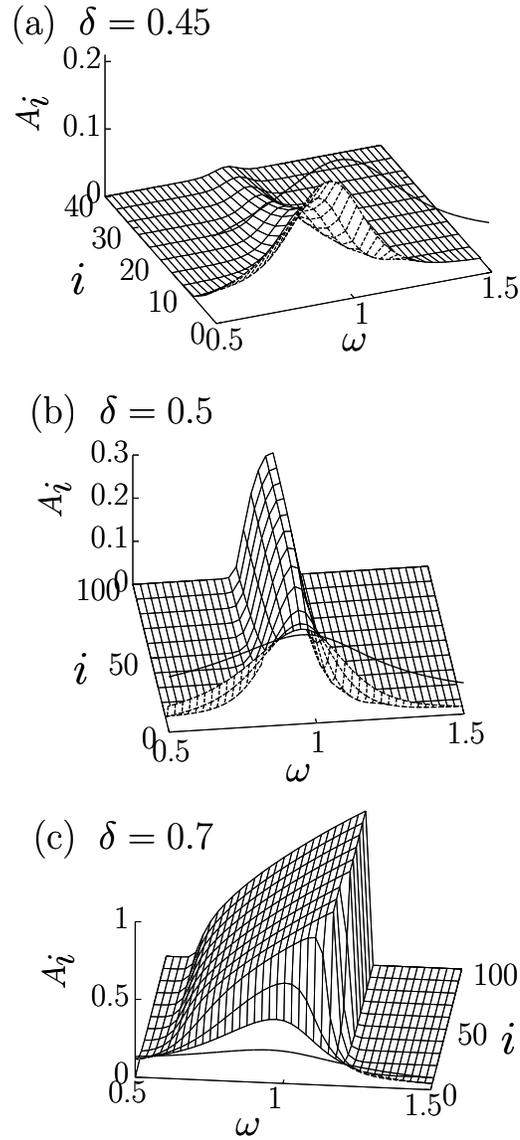}
\end{center}
\caption{Three-dimensional plot of $A_i$ versus $i$ and $\omega$ for three fixed values of $\delta$. $A_i$ is shown only for  some selected values of $i$. The thick curve represents $A_1$.}
\label{f5}
\end{figure}
The output signal damps out as it propagates through the coupled oscillators. When $\delta=0.5$ in Fig.~\ref{f4} we note that $A_{{\mathrm{L}}} \neq 0$ for a small range of values of $\omega$ implying undamped signal propagation and is evident in Fig.~\ref{f5}(b). The range of $\omega$ values for which undamped signal propagation occurs increases with increase in $\delta$. This is clearly seen in Fig.~\ref{f4} and \ref{f5}(b) and \ref{f5}(c).  In Fig.~\ref{f5}(c) where $\delta=0.7$ for first certain number of oscillators the response amplitude $A_i$ increases with $\omega$, reaches a maximum and then decreases smoothly. For distant oscillators after reaching a maximum value $A_i$ suddenly jumps to a lower value. When $\omega$ is increased from a smaller value to a higher value and then varied in the reverse direction $A_i$ follows the same path. $A_i$ is single-valued and there is no hysteresis phenomenon, however, jump phenomenon occurs. Next, we give an example of the case where $A_i$ shows hysteresis. Figure {\ref{f6}} illustrate the effect of one-way coupling on $A_{100}$ for three values of $\delta$ where $d=0.1$, $\omega_0^2=1$, $\beta=2$ and $f=0.2$. 
\begin{figure}[t]
\begin{center}
\includegraphics[width=0.35\linewidth]{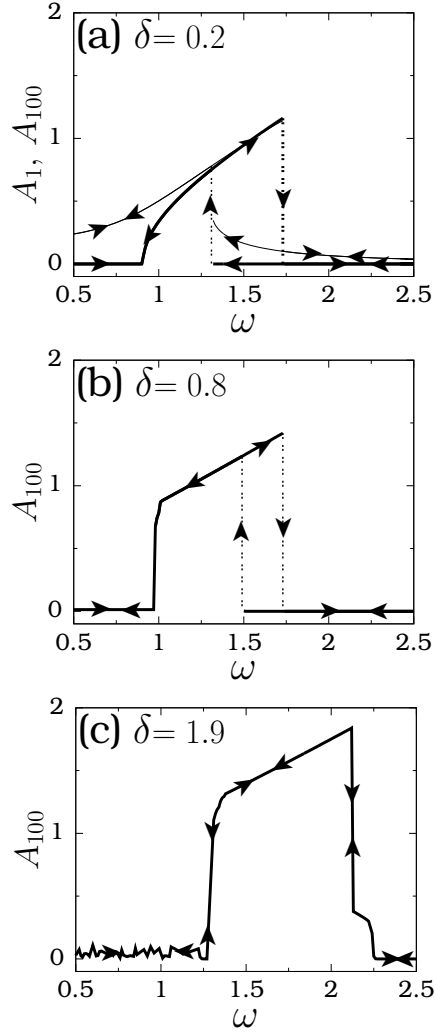}
\end{center}
\caption{The effect of the coupling strength $\delta$ on the hysteresis phenomenon. In the subplot (a) the thin and the thick curves represent $A_1$ and $A_{100}$ respectively. The arrows indicate the path followed by the response amplitude when $\omega$ is varied in the forward and reverse directions. The values of the parameters are $d=0.1$, $\omega_0^2=1$, $\beta=2$ and $f=0.2$. $A_1$ shown in the subplot (a) is independent of $\delta$ because the first oscillator is unaffected by the other oscillators in the system.}
\label{f6}
\end{figure}
The width of the hysteresis loop decreases with increase in $\delta$. In Fig.~\ref{f6}(c), for $\delta=1.9$, there is no hysteresis in $A_{100}$, however, we can clearly notice a sudden jump or a sharp variation in the value of $A_{100}$ at two values of $\omega$. Unidirectional coupling weakens and suppresses hysteresis.

\section{Coupled oscillators in two-dimensions}
In this section we extend our study to unidirectionally coupled oscillators in two-dimensions. Figure {\ref{f7}} depicts an example of $5 \times 5$ oscillators coupled unidirectionally in two-dimensions.  The first oscillator alone driven by the force $f\cos \omega t$. For simplicity we consider $n\times n$ oscillators. The equations of motion of the oscillators are given by
\begin{figure}[b]
\begin{center}
\includegraphics[width=0.38\linewidth]{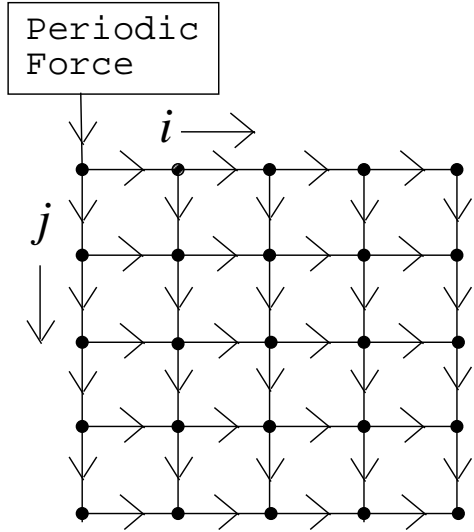}
\end{center}
\caption{A $5 \times 5$ coupled oscillators in two-dimensions. Periodic input signal $f\cos\omega t$ is fed into the first oscillator only. }
\label{f7}
\end{figure}
%
\begin{subequations}
\label{eq14} 
\begin{eqnarray} 
    \ddot{x}_{1,1} + d \dot{x}_{1,1} + \omega_0^2 x_{1,1} 
       + \beta x_{1,1}^3 
    & = & f \cos\omega t, \label{eq14a} \\
    \ddot{x}_{1,i} + d \dot{x}_{1,i} + \omega_0^2 x_{1,i}
       + \beta x_{1,i}^3 
    & = & \delta x_{1,i-1}, \label{eq14b} \\
    \ddot{x}_{j,1} + d \dot{x}_{j,1} + \omega_0^2 x_{j,1}
       + \beta x_{j,1}^3 
    & = & \delta x_{j-1,1}, \label{eq14c} \\
    \ddot{x}_{j,i} + d \dot{x}_{j,i} + \omega_0^2 x_{j,i}
       + \beta x_{j,i}^3 
    & = & \frac{1}{2} \delta \left( x_{j,i-1} + x_{j-1,i} \right),
            \label{eq14d}
\end{eqnarray}
\end{subequations}
where $j=2,3,\cdots, n$ and $i=2,3,\cdots, n$.

Seeking the periodic solution of Eqs.~(\ref{eq14}) in the form
\begin{subequations}
\label{eq15}
\begin{eqnarray} 
    x_{j,i}(t) = a_{j,i}(t) \cos\omega t
         + b_{j,i}(t) \sin \omega t, \label{eq15a} 
\end{eqnarray}
where
\begin{eqnarray}
    a_{j,i}(t) 
       =  A_{j,i}(t) \cos \theta_{j,i}(t), \quad
    b_{j,i}(t) 
       =  A_{j,i}(t) \sin \theta_{j,i}(t) ,
\end{eqnarray}
\end{subequations}
with $A_{j,i}^2 = a_{j,i}^2 + b_{j,i}^2$ we get the following set of equations:
\begin{subequations}
\label{eq16}
\begin{eqnarray}
    A_{j,i}^2 \left[ \omega_0^2 - \omega^2 
         + \frac{3}{4} \beta A_{j,i}^2 \right]^2
         + d^2 \omega^2 A_{j,i}^2 
     & = & F_{j,i}, \label{eq16a}
\end{eqnarray} 
where
\begin{eqnarray}
    F_{1,1}
      & = & f^2, 
            \quad F_{1,i} = \delta^2 A^2_{1,i-1}, \label{eq16b} \\
    F_{j,1}
      & = & \delta^2 A^2_{j-1,1}, \label{eq16c}  \\
    F_{j,i}
      & = & \frac{1}{4} \delta^2 \left( A^2_{j-1,i}
              + A^2_{j,i-1} \right) \label{eq16d} \nonumber \\
      &   &  \;\; + \frac{1}{2} \delta^2 A_{j-1,i} A_{j,i-1} 
              \cos (\theta_{j,i-1} - \theta_{j-1,i}) 
\end{eqnarray} 
with $i,j=1,2,\cdots, n$ and  
\begin{eqnarray}
   \theta_{1,i}
     & = & \theta_{1,i-1} + \tan^{-1} \left(
             \frac{d \omega}{\omega_0^2 - \omega^2
               + \frac{3}{4}\beta A_{1,i}^2 } \right),
                 \label{eq16e} \\
   \theta_{j,1}
     & = & \theta_{j-1,1} + \tan^{-1} \left( 
            \frac{d \omega}{\omega_0^2 - \omega^2 
              + \frac{3}{4} \beta A_{j,1}^2 } \right),
                 \label{eq16f} \\
    d \omega A_{j,i}
     & = & \frac{1}{2} \delta A_{j-1,i}
             \sin \left( \theta_{j,i} - \theta_{j-1,i} \right)
              + \frac{1}{2} \delta A_{j,i-1} 
              \sin \left( \theta_{j,i} - \theta_{j,i-1} \right),
                \label{eq16g}
\end{eqnarray} 
\end{subequations}
where $\theta_{1,0}=\theta_{0,1}=0$.

For the oscillators numbered as $(j,1)$ and $(1,i)$ the equations for $A_{j,1}$ and $A_{1,i}$ are independent of $\theta_{j,1}$ and $ \theta_{1,i}$ respectively and are the solutions of the cubic equations of the square of the amplitudes given by the Eqs.~(\ref{eq16a}-{c}). $\theta_{1,i}$ and $\theta_{j,1}$ are given by the Eqs.~(\ref{eq16e}) and (\ref{eq16f}) respectively. For the oscillators numbered as $(j,i)$, $j,i \neq 1$ the equations for $A_{j,i}$ depend on $\theta_{j,i-1}$ and $\theta_{j-1,i}$ but independent of $\theta_{j,i}$. Therefore $A_{j,i}$'s are the solutions of the Eqs.~(\ref{eq16a}) and (\ref{eq16d}). Then we compute $\theta_{j,i}$ from Eq.~(\ref{eq16g}) using Newton--Raphson method.

First, we point out the results obtained without solving the above set of Eqs.~(\ref{eq16}). The set of $n$ oscillators in the first-row and first-column in Fig.~\ref{f7}, that is, the oscillators numbered as $(j,i) \rightarrow (1,i)$, $i=1,2,\cdots, n$ and $(j,i) \rightarrow (j,1)$, $j=1,2,\cdots, n$ essentially form two identical set of coupled oscillators in one-dimension. We note that 
\begin{eqnarray} 
\label{eq17}
   A_{1,k} = A_{k,1} \;\; {\mathrm{and}} \;\;  
       \theta_{1,k} = \theta_{k,1}, \;\; k=2,3,\cdots,n.
\end{eqnarray} 
Suppose
\begin{subequations}
\label{eq18}
\begin{eqnarray}
    A_{1,i} & \neq & A_{{\mathrm{L}}}, \;\; i=1,2,\cdots, k
                 \leq n \label{eq18a} \\
   A_{1,i} & = & A_{{\mathrm{L}}}, \;\; i=k+1,k+2,\cdots, n. \label{eq18b}
\end{eqnarray}
\end{subequations}
Then because the coupling term of $(j,i)$th oscillator is $\delta \left( x_{j,i-1}+x_{j-1,i}\right)/2$ we realize that for each $l$ where $l=1,2,\cdots, k$ all the
\begin{equation}
\label{eq19}
   A_{j,l+1-j}, \quad j=1,2,\cdots, k
\end{equation}
are same while all the other $A_{j,i}$'s become $A_{{\mathrm{L}}}$. For example, if $A_{1,i} \neq A_{{\mathrm{L}}}$ for $i=1,2,3$ and $A_{1,i}=A_{{\mathrm{L}}}$ for $i=4,5,\cdots, n$ then 
\label{eq20}
\begin{eqnarray}
   A_{1,2} = A_{2,1} \neq A_{{\mathrm{L}}}, \quad
       A_{1,3} = A_{2,2} = A_{3,1} \neq A_{{\mathrm{L}}} \label{eq20}
\end{eqnarray}
while all other $A_{j,i}$'s are $A_{{\mathrm{L}}}$. This is depicted in Fig.~\ref{f8} where the oscillators with same amplitude but different from $A_{{\mathrm{L}}}$ are connected by lines while for the oscillators represented
\begin{figure}[b]
\begin{center}
\includegraphics[width=0.36\linewidth]{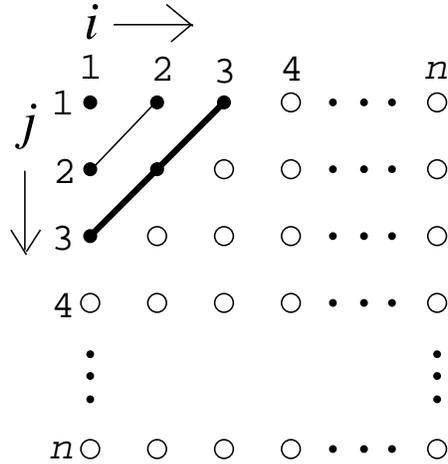}
\end{center}
\caption{For the case of $A_{1,i} \neq A_{{\mathrm{L}}}$ for $i=1,2,3$ and $A_{1,i} = A_{{\mathrm{L}}}$ for $i=4,5,\cdots,n$ the oscillators represented by solid circles connected by same lines have the same amplitudes but not equal to $A_{{\mathrm{L}}}$ while the oscillators represented by open circles have the same amplitude $A_{{\mathrm{L}}}$.}
\label{f8}
\end{figure}
 by open circles $A_{j,i} = A_{{\mathrm{L}}}$. Because $A_{12} \neq A_{13}$ the two set of oscillators numbered as $(j,i)\rightarrow(1,2)$ and $(2,1)$ and $(1,3)$, $(2,2)$ and $(3,1)$ are joined by lines with different thickness. $A_{{\mathrm{L}}}$ of the system (\ref{eq14}) is same as that of the system (\ref{eq1}).

We fix the values of the parameters as $d=0.5$, $\omega_0^2=1$, $\beta=1$ and $f=0.1$. For $n \times n$, with $n=100$, coupled oscillators we calculate $A_{j,i}$ by solving the Eqs.~(\ref{eq16}) and then compute the average gain of the response amplitude given by
\begin{equation}
\label{eq21}
   \overline{G} = \frac{1}{n^2 A_{1,1}} \sum^n_{j=1} 
          \sum^n_{i=1} A_{j,i} .
\end{equation}
Figure {\ref{f9}(a)} presents $\overline{G}$ as a function of the coupling strength $\delta$ and the frequency $\omega$ of the driving force. In Fig.~\ref{f9}(b) the result for the coupled oscillators in one-dimension is shown for comparison. Both Figs.~\ref{f9}(a) and (b) are almost identical. 
\begin{figure}[b]
\begin{center}
\includegraphics[width=0.53\linewidth]{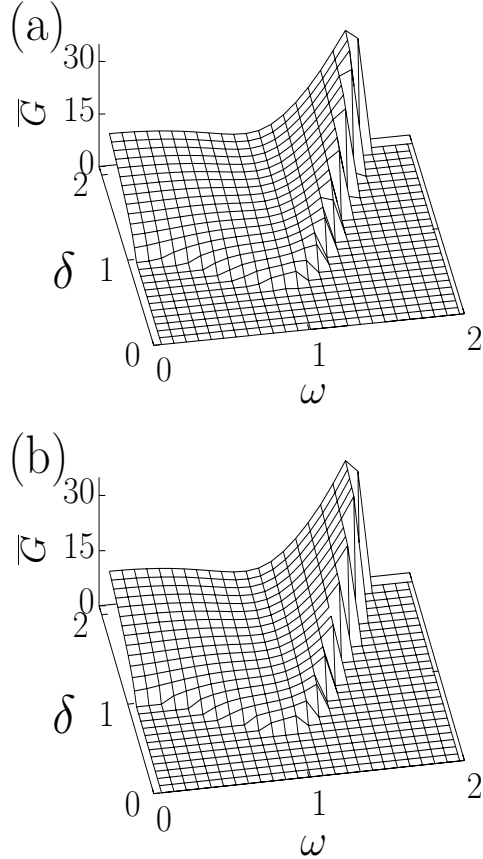}
\end{center}
\caption{{Dependence of average gain given by Eq.~(\ref{eq21}) on  $\delta$ and $\omega$ for the coupled oscillators in (a) two-dimensions and (b) one-dimension with the first oscillator alone driven by the periodic force.}}
\label{f9}
\end{figure}
Figure \ref{f10} illustrates the variation of the response amplitude $A_{j,i}$ with $(j,i)$ for three fixed values of $\delta$ with $\omega=0.7$. For $\delta=0.6 < \delta_{{\mathrm{u}}}(=f/A_1)$ the $A_{j,i}$ decays to zero with increase in $j$ and $i$. Enhanced undamped signal propagation with $A_{j,i}$ evolving to a saturation along all the directions can be clearly seen in Figs.~\ref{f10}(b) and (c) for $\delta=0.7$ and $1.5 >\delta_{{\mathrm{u}}}$ .

\begin{figure}[t]
\begin{center}
\includegraphics[width=0.48\linewidth]{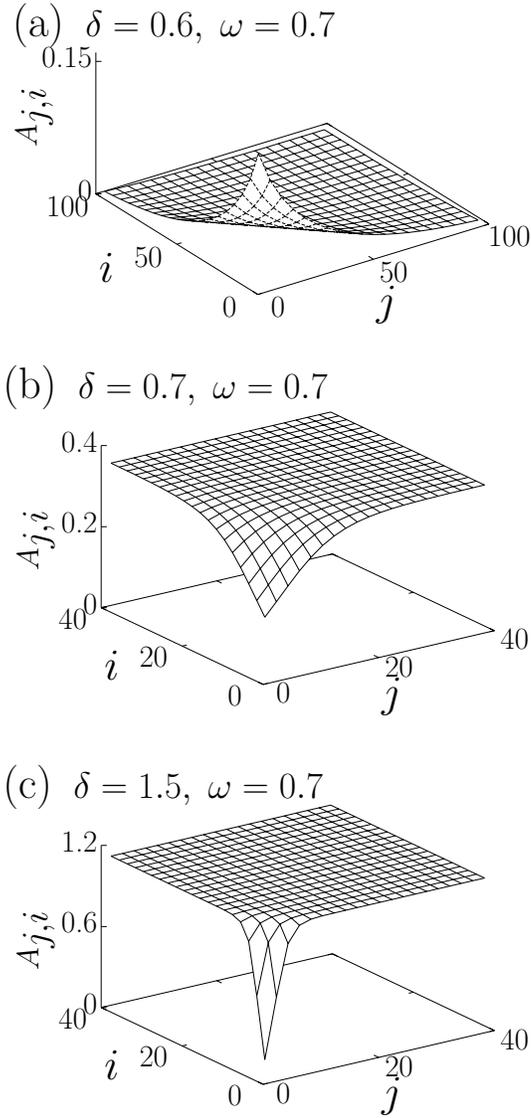}
\end{center}
\caption{Three-dimensional plot of the response amplitude $A_{j,i}$ of the system (\ref{eq14}) as a function of $i$ and $j$ for three fixed values of the coupling strength $\delta$. The values of the other parameters are $d=0.5$, $\omega_0^2=1$, $\beta=1$, $f=0.1$ and $\omega=0.7$.}
\label{f10}
\end{figure}
%
\section{Randomly coupled and forced oscillators in two-dimensions}
In this section we consider two more configurations of coupled oscillators in two-dimensions. In the previous section we analysed the network where the first oscillator alone driven by a periodic force while the other oscillators are coupled unidirectionally. Now, we consider the following two types of coupled oscillators:

\begin{enumerate}
\item 
{\bf{Randomly coupled oscillators}}

All the oscillators are driven by the periodic force $f \cos\omega t$. For $n\times n $ oscillators the total number of connections is $2n(n-1)$. Out of these only $N_{{\mathrm{C}}}\%$ of total connections chosen randomly are switched-on. Figure \ref{f11} shows an example of the network with $50\%$ of the total connections are set on.
\item 
{\bf{Randomly forced oscillators}}

All the oscillators except the first oscillator are coupled unidirectionally. However, periodic force is applied to only $N_{{\mathrm{C}}}\%$ of total number $(n^2)$ of oscillators chosen randomly.
\end{enumerate}
%

\begin{figure}[b]
\begin{center}
\includegraphics[width=0.38\linewidth]{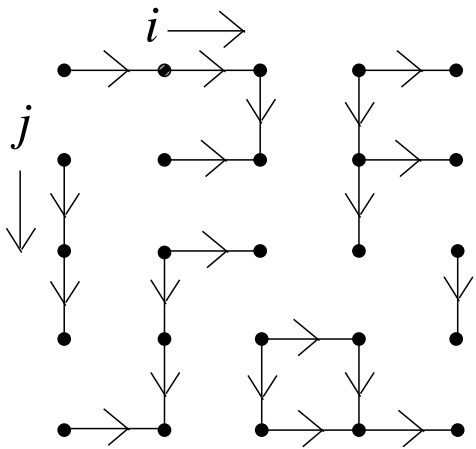}
\end{center}
\caption{ An example of $5 \times 5$ two-dimensional oscillators with $50\%$ of the one-way couplings chosen randomly are set on. The other connections are switched-off. All the oscillators are driven by the periodic force $f\cos \omega t$.}
\label{f11}
\end{figure}

With reference to the above types of coupled network  systems it is worth to mention the relevance of such network configurations in neuroscience.  In the nervous system the structure permitting to transmit an information from one neuron to another is the synapse.  In neuroscience synaptic plasticity is the ability of the nervous system to modify the neural network to realize an enhanced or depressed efficacy of synaptic transmission and experimentally several forms of synaptic plasticity  have been measured.  Response of a neural network can be modified by changing the value of coupling  strength or establishing new connections or breaking certain connections in the network {\cite{r42,r43}}.  Such  changes in a neural circuity lead to recovery of functions after brain injury and mental illness.  Synaptic plasticity is found to play important roles in fast network configuration and fast adaptation in free operand learning. 

Interestingly, for the randomly coupled oscillators and randomly forced oscillators also the theoretical approach used in the previous sections can be employed. The equation of motion of the randomly coupled oscillators is 
\begin{subequations}
\label{eq22}
\begin{eqnarray} 
   \ddot{x}_{1,1} + d \dot{x}_{1,1} + \omega_0^2 x_{1,1}
           + \beta x_{1,1}^3
    & = & f \cos\omega t,   \label{eq22a}  \\
   \ddot{x}_{1,i} + d \dot{x}_{1,i} + \omega_0^2 x_{1,i}
           + \beta x_{1,i}^3 
    & = & \delta_{1,i-1} x_{1,i-1} + f \cos\omega t, \label{eq22b}  \\
   \ddot{x}_{j,1} + d \dot{x}_{j,1} + \omega_0^2 x_{j,1}
           + \beta x_{j,1}^3 
    & = & \delta_{j-1,1} x_{j-1,1} + f \cos\omega t,  \label{eq22c}  \\
   \ddot{x}_{j,i} + d \dot{x}_{j,i} + \omega_0^2 x_{j,i}
           + \beta x_{j,i}^3 
    & = & \frac{1}{2} \left( \delta_{j,i-1} x_{j,i-1}
           + \delta_{j-1,i} x_{j-1,i} \right)
             + f \cos\omega t .  \label{eq22d}
\end{eqnarray}
\end{subequations}
In Eqs.~(\ref{eq22}) $\delta_{j,i-1}=\delta$, a constant if the one-way coupling between the oscillators $(j,i)$ and $(j,i-1)$ is set on otherwise $\delta_{j,i-1}=0$. Similarly, we can set the value $\delta$ or $0$ to the quantities $\delta_{k,l}$ in Eqs.~(\ref{eq22}). For the system (\ref{eq22}) the equations for the response amplitudes $A_{j,i}$ are given by Eq.~(\ref{eq16}a) with 
\begin{subequations}
\label{eq23}
\begin{eqnarray}
    F_{1,1} 
      & = &  f^2, \quad  F_{1,i} = \delta^2_{1,i-1} A^2_{1,i-1}
             + f^2, \label{eq23a}  \\
    F_{j,1}
      & = &  \delta^2_{j-1,1} A^2_{j-1,1} + f^2, \label{eq23b}  \\
    F_{j,i}
      & = &  \frac{1}{4} \left( \delta^2_{j-1,i} A^2_{j-1,i}
              + \delta^2_{j,i-1} A^2_{j,i-1} \right)
               \nonumber \\
      &   &  + \frac{1}{2} \delta_{j-1,i} \delta_{j,i-1}
                 A_{j-1,i} A_{j,i-1} 
                  \cos \left( \theta_{j,i-1} 
                 - \theta_{j-1,i}  \right),  \label{eq23c}
\end{eqnarray}
\end{subequations}
where $j=2,3,\cdots, n$, $i=2,3,\cdots, n$. $\theta_{j,i}$ can be obtained from 
\begin{subequations}
\label{eq24}
\begin{eqnarray}
   d \omega A_{j,i} 
     & = &  \frac{1}{2} \delta_{j-1,i} A_{j-1,i}
            \sin \left( \theta_{j,i} - \theta_{j-1,i} \right)
               \nonumber \\   
     &   &  + \frac{1}{2} \delta_{j,i-1} A_{j,i-1} 
              \sin \left( \theta_{j,i} - \theta_{j,i-1} \right)
               + f \sin \theta_{j,i} . 
\end{eqnarray} 
\end{subequations}
where $\theta_{1,1}$ is given by Eq.~(\ref{eq16f}) while $\theta_{0,i}$ and $\theta_{j,0}$ are all zero because the starting values of $i$ and $j$ are $1$.

For a fixed $N_{{\mathrm{C}}}$ and for one realization of the coupled oscillators we calculate $A_{j,i}$ from Eqs.~(\ref{eq16a}) and (\ref{eq23}) and then compute an average gain factor $\overline{G}$ given by Eq.~(\ref{eq21}) of the response amplitudes. Repeating this procedure for different realizations of coupled oscillators we obtain corresponding $\overline{G}$. Then we compute the average value of $\overline{G}$ and denote it as $\langle \overline{G} \rangle$.  We fix $d=0.5$, $\omega_0^2=1$, $\beta=1$, $f=0.1$ and $n=100$.   In Fig.~\ref{f12} we plot $\langle \overline{G} \rangle$ as a function of $N_{{\mathrm{R}}}$, the number of different realizations of coupled oscillators, for $N_{{\mathrm{C}}}=40\%$, $\delta=2$ and $\omega=1.28$.  $\langle \overline{G} \rangle$ increases with $N_{{\mathrm{R}}}$ and becomes almost constant for $N_{{\mathrm{R}}}>70$. For $N_{{\mathrm{R}}}=100$ we found $\langle \overline{G} \rangle=4.374$ with standard deviation $0.089$.  In the case of $N_{{\mathrm{R}}}=200$, $\langle \overline{G} \rangle=4.370$ with standard deviation $0.089$.  In our further analysis we fix $N_{{\mathrm{R}}}=100$.  Though $N_{{\mathrm{R}}}=100$ is quite small, for each realization of coupled oscillators $\overline{G}$ is calculated averaging over $10^4$ oscillators which is quite large.  

\begin{figure}[b]
\begin{center}
\includegraphics[width=0.4\linewidth]{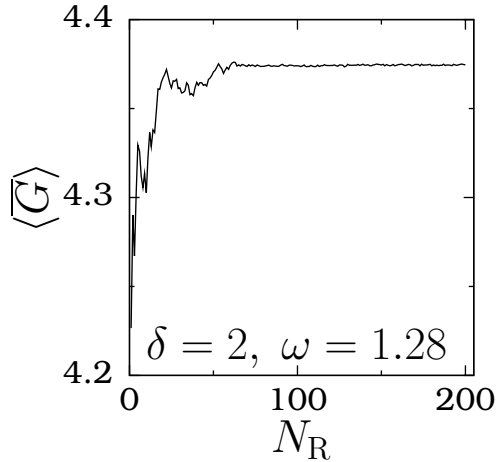}
\end{center}
\caption{Variation of $\langle \overline{G} \rangle$ as a function of $N_{{\mathrm{R}}}$ for the randomly coupled $100 \times 100$ oscillators in two-dimensions with $N_{{\mathrm{C}}}=40\%$.   The values of the parameters are $d=0.5$, $\omega_0^2=1$, $\beta=1$, $f=0.1$, $\delta=2$ and $\omega=1.28$. }
\label{f12}
\end{figure}

Figure \ref{f13} shows $\langle \overline{G}\rangle$ versus $\delta$ and $\omega$ for four values of $N_{{\mathrm{C}}}$. In the undamped signal propagation regions of $(\delta,\omega)$ the quantity $\langle \overline{G}\rangle$ increases with increase in $N_{{\mathrm{C}}}$. 
\begin{figure}[t]
\begin{center}
\includegraphics[width=0.38\linewidth]{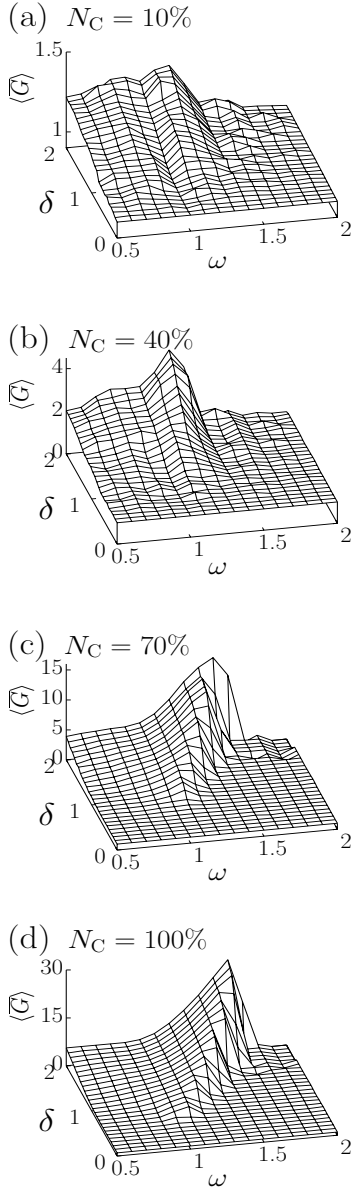}
\end{center}
\caption{Plot of the average gain $\langle \overline{G} \rangle$ as a function of the control parameters $\delta$ and $\omega$ for the coupled oscillators in two-dimensions with one-way coupling is made for $10\%$, $40\%$, $70\%$ and $100\%$ of total number of oscillators. For each oscillator connectivity is set either on or off randomly. }
\label{f13}
\end{figure}
Compare Fig.~\ref{f13}(d) where $N_{{\mathrm{C}}}=100\%$ (that is, all the oscillators are coupled and driven by the periodic force $f \cos\omega t$) with  Fig.~\ref{f9}(a) corresponding to the network with all the oscillators are coupled but first oscillator alone driven by the periodic force. For a range of values of $\delta$ and $\omega$ the values of $\langle \overline{G}\rangle$ in Fig.~\ref{f13}(d) is lower than that in Fig.~\ref{f9}(a). The important result is that driving all the oscillators by the periodic force $f\cos \omega t$ not increases $\langle \overline{G}\rangle$ compared to the case of the first oscillator alone driven. To illustrate the effect of driving all the oscillators in Fig.~\ref{f14}a we plot $A_{j,i}$ versus $j$ and $i$ for $N_{{\mathrm{C}}}=100\%$. This figure can be compared with the Fig.~\ref{f10}(c) where only the first oscillator is driven. 
All the oscillators exhibit periodic motion with period $T=2 \pi / \omega$.  Therefore, the one-way coupling term can be viewed as a second external periodic force.  Figure \ref{f14}b shows the variation of $\theta_{j,i}$ with $j$ and $i$.  $\theta_{j,i}$ varies with $j$ and $i$.   This is clear from Fig.~\ref{f14}c.  The motion of the oscillators are periodic (with different amplitude)  but they are not synchronized.  That is, the  irregular variation of  $A_{j,i}$ and reduction in their values is due to the phase difference between the applied periodic force and the coupling term in Eqs.~(\ref{eq22}).

\begin{figure}[t]
\begin{center}
\includegraphics[width=0.8\linewidth]{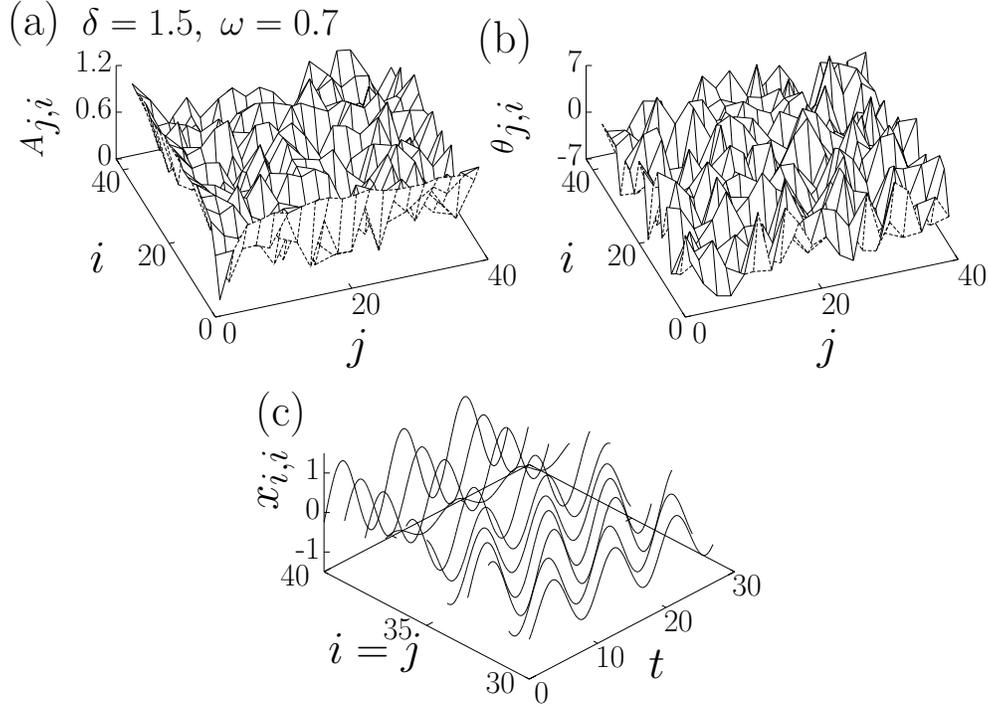}
\end{center}
\caption{(a) $A_{j,i}$ versus $j$ and $i$ and (b) $\theta_{j,i}$ versus $j$ and $i$ for the coupled oscillators with all the oscillators driven by the periodic force.  (c) $x_{i,i}$ versus $t$ for $i=30,31,\cdots ,40$. Here $N_{{\mathrm{C}}}=100\%$. }
\label{f14}
\end{figure}
%
\begin{figure}[t]
\begin{center}
\includegraphics[width=0.55\linewidth]{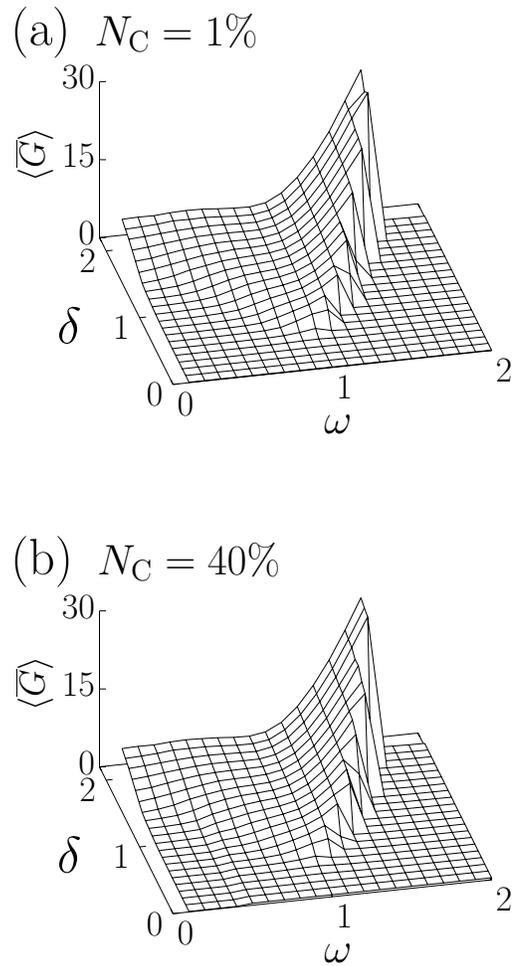}
\end{center}
\caption{Variation of gain factor response of the unidirectionally coupled oscillators with $N_{{\mathrm{C}}}\%$ of total number of oscillators chosen randomly driven by the periodic force.}
\label{f15}
\end{figure}

Finally, we consider the randomly forced coupled oscillators. For this system also we apply the theoretical method used for the system (\ref{eq22}) and compute $\langle \overline{G}\rangle$. Figure \ref{f15} displays $\langle \overline{G}\rangle$ versus $\delta$ and $\omega$ for $N_{{\mathrm{C}}}=1\%$ and $40\%$. We do not observe any appreciable difference between the cases of $N_{{\mathrm{C}}}=1\%$ and $40\%$. 

\section{Conclusions}
In conclusion, in the unidirectionally coupled Duffing oscillators in both one and two-dimensions with the first oscillator alone driven by a periodic force, we have shown the occurrence of enhanced signal propagation. Above a critical value of the coupling strength the response amplitude increases with increase in the number of oscillators and reaches a saturation.  That is, there exit a critical number of oscillators to have a maximum response and further increase in the number of oscillators is ineffective in enhancing the response. The applied perturbation method enables us to  predict (i) the response amplitude of each oscillator, (ii) the saturation value of the response amplitude and (iii) the critical value of the coupling strength above which undamped signal propagation occurs. Increase in the coupling strength leads to decrease in the width of the hysteresis and above a critical value of it hysteresis is suppressed, however, jump phenomenon persists. We have shown the applicability of the perturbation method for the cases of the coupled oscillators with fraction of total number of oscillators chosen for coupling and forcing.

 \vskip 5pt
\noindent{\bf{Acknowledgment}}
 \vskip 5pt
S.~Rajamani expresses her gratitude to University Grants Commission (U.G.C.), India for financial support in the form of U.G.C. meritorious fellowship.
\end{document}